\def\RELEASE{0}  %
\def\ANON{0}     %
\def\SQUEEZE{0}  %

\documentclass[sigplan,10pt,screen]{acmart} %
\PassOptionsToPackage{hyphens}{url}\usepackage{hyperref}

\renewcommand\footnotetextcopyrightpermission[1]{} %
\usepackage[switch]{lineno}

\usepackage[noend]{algpseudocode}
\usepackage{algorithmicx}
\usepackage{booktabs}
\usepackage{comment}
\usepackage{graphicx}
\usepackage{listings}
\usepackage{multirow}
\usepackage{xcolor}
\usepackage{xspace}
\usepackage[varqu]{zi4}
\usepackage[T1]{fontenc}
\usepackage{tikz}
\usepackage{gnuplot-lua-tikz}
\usepackage{lipsum}
\usepackage{paralist}

\microtypecontext{spacing=nonfrench}

\hypersetup{
  pdflang={en},
  pdfdisplaydoctitle}

\definecolor[named]{OurPurple}{cmyk}{0.55,1,0,0.15}
\definecolor[named]{OurDarkBlue}{cmyk}{1,0.58,0,0.21}

\definecolor{purp_cb}{HTML}{33159E}
\definecolor{blue_cb}{HTML}{6D8EF7}
\definecolor{gree_cb}{HTML}{61D19B}

\hypersetup{
  colorlinks,
  linkcolor=OurPurple,
  citecolor=OurPurple,
  urlcolor=OurDarkBlue,
  filecolor=OurDarkBlue}

\definecolor{purp_cb}{cmyk}{0.677,0.867,0.000,0.380} %
\definecolor{blue_cb}{cmyk}{0.559,0.425,0.000,0.031} %
\definecolor{gree_cb}{cmyk}{0.536,0.000,0.258,0.180} %
\definecolor{oran_cb}{cmyk}{0.000,0.183,0.554,0.059} %
\definecolor{yell_cb}{cmyk}{0.012,0.000,0.383,0.047} %

\if \SQUEEZE 1
\setlist[itemize]{
  leftmargin=*,
  itemsep=2pt,
  topsep=2pt}

\fi
\usepackage{relsize}

\def\Snospace~{\S{}}

\if \RELEASE 1
  \def\NOTES{0}
\else
  \def\NOTES{1}
\fi
\if \NOTES 1
  \newcommand{\XXX}[1]{{\color{red}{XXX {#1}}}}
  \newcommand{\antoine}[1]{{\color{teal}{[\textbf{AK:} {#1}]}}}
  
  \newcommand{\vaas}[1]{{\color{orange}{[\textbf{VA:} {#1}]}}}
  \newcommand{\todo}[1]{{\color{blue}{TODO: {#1}}}}
\else
  \newcommand{\XXX}[1]{}
  \newcommand{\antoine}[1]{}
  \newcommand{\vaas}[1]{}
  \newcommand{\authorC}[1]{}
  \newcommand{\todo}[1]{}
\fi

\newcommand{\figscale}{0.65}

\if \ANON 1
  \newcommand{\sys}{Consigliere\xspace}
\else
  \newcommand{\sys}{LaissezCloud\xspace}
\fi

\lstdefinelanguage{Go}{
  keywords={break,case,chan,const,continue,default,defer,else,fallthrough,for,
    func,go,goto,if,import,interface,map,package,range,return,select,struct,
    switch,type,var},
  sensitive=true,
  morecomment=[l]{//},
  morecomment=[s]{/*}{*/},
  morestring=[b]",
}
\title{ \sys: Continuous Resource Renegotiation for the Public Cloud}

\if \ANON 1
  \author{Paper \#911}
\else
  \author{
    Tejas Harith, Antoine Kaufmann\\
    MPI-SWS
    }
\fi
\date{}

\begin{abstract}
Public clouds increasingly expose heterogeneous hardware, but their allocation
interface remains built around rigid on-\-demand and spot service classes.
This makes it hard to satisfy time-varying tenant objectives and operator
constraints in oversubscribed, heterogeneous clusters without exposing
internal application or infrastructure state.
We present \sys, a cloud resource management platform for continuous
re-negotiation of running allocations.
Unlike spot instances, which use launch-time bids and unilateral preemption,
\sys keeps allocations continuously contestable during execution:
tenants and operators update bids online, and a running tenant
keeps a resource only as long as its bid exceeds competing demand.
Pricing serves both as a narrow waist and as an incentive-alignment mechanism
between mutually untrusted participants: tenants express utility through bids,
while operators price in power, cooling, or carbon constraints without exposing
internal telemetry.
Across a diverse set of accelerator workloads, \sys reduces performance
degradation under contention by 8--23\% versus on-demand and spot baselines, and
scales to clusters of at least 10{,}000 nodes.
\end{abstract}
 
\begin{document}
\pagestyle{plain}
\maketitle

\section{Introduction}%
\label{sec:intro}

Today's public-cloud contract is static: tenants choose hardware and pricing
mode at launch, and that decision governs execution until the instance is
released or preempted.
This does not match modern cloud realities, where specialized hardware,
oversubscribed capacity, and changing infrastructure conditions make
the value of a running allocation inherently
time-varying~\cite{company:crusoe,company:lambdalabs}.
Crucially, this heterogeneity is not limited to hardware type: even two
instances of the same type can differ sharply in value depending on their
locality to a tenant's existing allocation or their position in the cloud's
failure-domain hierarchy.
On-demand fixes price and resource choice too early.
Spot instances offer operators flexibility, but tenants can only express
launch-time bids and must tolerate unilateral and unexpected preemption.
Neither model supports revisiting a running allocation in a coordinated way,
as tenant utility, competing demand, and infrastructure conditions change.
Current clouds lock in suboptimal placements, degrade performance under
contention, and leave operators with limited control \& flexiblity~\cite{wu:cantbelate,yang:skypilot}.

Improving allocations requires coordination across tenants and operator.
That is practical in a single administrative
domain~\cite{eva,stratus,cilantro,htas}, but the public cloud adds a hard trust
boundary.
Tenants do not want to reveal workload internals, to operators or other tenants.
Operators also do not want to expose telemetry, pricing logic, or infrastructure
constraints.
What is missing is an interface that enables coordination and reallocation
without shared control or detailed transparency.

Existing work addresses only parts of this problem.
Cluster schedulers and infrastructure managers improve efficiency inside
a single administrative domain \cite{stojkovic:tapas,stojkovic:dynamollm,
ecovisor,faro}, especially with broad visibility into workloads and
infrastructure state \cite{eva,stratus,cilantro,htas}.
Cloud application frameworks, in contrast, adapt to today's on-demand and
spot interfaces rather than changing the interface itself
\cite{wu:cantbelate,yang:skypilot}.
Of the few systems that unify multiple dynamic tenants
\cite{vuppalapati:karma,grandl:carbyne,ghodsi:drf,chaudhary:gandivafair,xchange,10.1145/3620665.3640375,sf_compute,themis,shockwave},
none provides a mechanism for continuous reallocation across mutually untrusted
tenants and
operators.

Our key insight is that price can serve as the common interface among
tenants and the cloud without forcing any participant to expose internals.
Because price is already the cloud's common contract, tenants can express the
current utility of a resource through bids while operators express
infrastructure pressure through updated prices.
Pricing thus acts both as a narrow waist and as an
incentive-alignment mechanism across mutually untrusted participants,
including competing tenants.
This lets running allocations be re-negotiated continuously rather than
fixed at launch.

We realize this price-based interface in \sys as a live market for individual
compute resources, using a topology-aware matching engine and operator-side
\textsc{InfraMaps}.
The matching engine captures locality and failure-domain structure, letting tenants
value resources not just by type, but by where they sit relative to an
existing allocation.
Tenant \textsc{EconAdapters} translate application utility
and reconfiguration cost into active bids, limits, and relinquishment decisions,
while
operator \textsc{InfraMaps} translate telemetry and policy into
resource-specific price adjustments.
When demand or operator pressure changes, tenants can rebid, migrate, or
relinquish resources rather than waiting for allocations to end or for
unilateral preemption.

In our evaluation, \sys reduces performance degradation under contention by
8--23\% relative to on-demand and spot.
For topology-sensitive workloads, \sys's locality-aware bidding enables
resource exchanges that can nearly double training performance.
It also provides operators a lever to steer demand using price signals, e.g.
to shift load away from power-constrained domains without exposing telemetry.
Despite increased dynamic interaction compared to today's cloud,
\sys scales to clusters of at least 10{,}000 nodes.
Continuous negotiation improves efficiency, expands operator control,
and is practical at scale.

\noindent
This paper makes the following contributions:

\smallskip
\begin{compactitem}[\labelitemi]
  \item We introduce a public-cloud interface for continuous negotiation of
    running allocations via price.
    
  \item We design \sys, a topology-aware market that keeps arbitration
    centralized while leaving valuation and control policies local to tenants
    and operators.
  
  \item We evaluate \sys on heterogeneous workloads under
    contention and operator-side infrastructure constraints.
\end{compactitem}

\section{Motivation}
\label{sec:motivation}

\subsection{Why Static Cloud Resource Contracts Fail}
Once public clouds expose scarce, specialized hardware in heterogeneous,
oversubscribed clusters constrained by shared infrastructure such as power,
cooling, and network capacity
\cite{10.1145/2503210.2504566,aws-scalable-hwagnostic-inference}, a one-time
launch decision is no longer adequate.
Rather than a single fungible pool, modern clouds are increasingly collections
of finite, non-fungible resource pools whose instantaneous value depends on
both hardware type and placement.
Even two resources of the same instance type can differ sharply in value: a
second GPU in the same host or NVLink domain may be far more valuable to a
training job than the same GPU in another rack, while availability-sensitive
tenants may prefer the opposite.
As a result, operators must often choose not just which resource type to
reclaim or reassign, but which specific resource instance.

Operators need ways to react to changing demand and infrastructure
conditions, but spot-style preemption is only a coarse lever: it creates
headroom only when tenants are willing to accept interruption.
Even when enough spot capacity exists, the operator may need to free any $N$
resources from a feasible set, but spot reveals only general willingness to be
preempted, not the current inconvenience of interruption.
Since tenants' interruption costs differ across applications and over time, the
operator is left to make reclaim decisions without knowing which tenant would
be the best candidate.%

For many modern workloads, the value of a resource is time-varying: it depends
on the application's current phase, the cost of reconfiguration, and evolving
application demand.
A training job may be cheap to interrupt just after a checkpoint but expensive
to move in the middle of an iteration.
Current cloud interfaces expose only a fixed launch-time choice or abrupt
preemption, even though both tenant utility and operator constraints evolve
throughout execution.

Despite the public-cloud trust boundary, improving these allocations still
requires coordination across tenants and the
operator~\cite{10.1145/3620665.3640375,222611}.
What is missing is a thin interface that supports continuous reallocation and
operator steering without requiring either side to expose internal state.

\begin{figure}%
\centering
\footnotesize
\begin{tabular}{cccll}
    \toprule
    \multicolumn{1}{c}{\textbf{Accelerator}} &
        \multicolumn{2}{c}{\textbf{Exec. Time}}  &
        \multicolumn{2}{c}{\textbf{Exec. Cost}} \\
    & \textcolor{blue_cb}{\textbf{A}} &
      \textcolor{gree_cb}{\textbf{B}} &
      \multicolumn{1}{c}{\textcolor{blue_cb}{\textbf{A}}} &
      \multicolumn{1}{c}{\textcolor{gree_cb}{\textbf{B}}} \\
    \midrule
    HW1  & \textcolor{blue_cb}{0.35}      &
        \textcolor{gree_cb}{\textbf{0.23}}    &
        \textcolor{blue_cb}{\textbf{\$0.21}}  &
        \textcolor{gree_cb}{\textbf{\$0.14}} \\
    HW2  & \textcolor{blue_cb}{0.51}      &
        \textcolor{gree_cb}{0.32}    &
        \textcolor{blue_cb}{\$0.25} (+15\%)    &
        \textcolor{gree_cb}{\$0.15} (+8\%) \\
    HW3  & \textcolor{blue_cb}{\textbf{0.30}}      &
        --      &
        \textcolor{blue_cb}{\$0.24} (+13\%)    &
        -- \\
    \bottomrule
\end{tabular}

\includegraphics[scale=\figscale]{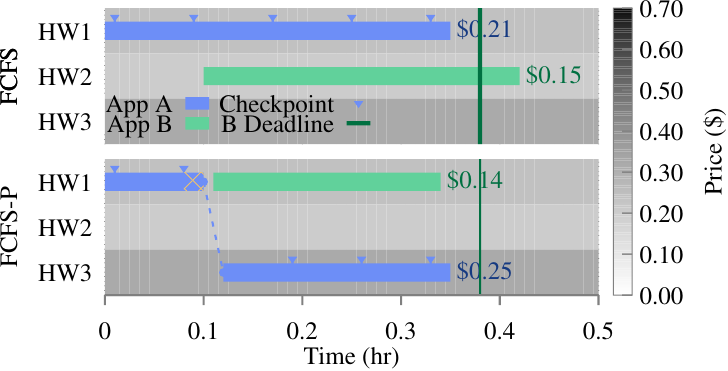}%
\caption{FCFS: App A acquires HW1 first, forcing App B, dispatched at 0.1, to take
its second-ranked hardware. FCFS-P: App A acquires HW1 first, but App B, dispatched at 0.1, preempts
App A, interrupting an epoch between checkpoints.}%
\label{fig:motiv_fcfs_fcfsp}%
\end{figure}

\subsection{An Illustrative Cloud Example}
\label{sec:motiv_ex}
To make this concrete, consider two applications running on a small cloud with
three machines of different hardware types.
App A is a checkpointed training task and App B is a deadline-sensitive batch
analytics task.
We consider a cloud objective of high utilization and tenant objectives of low
cost per completion.
\autoref{fig:motiv_fcfs_fcfsp} shows execution time and cost for the applications
across hardware types.
The same argument applies when multiple instances of the same type differ only
in locality or failure-domain position: one placement may be substantially
more valuable to a tenant than another.

With on-demand, whichever application acquires HW1 first keeps it for the
duration of execution, so if A arrives first, B blocks even though HW1 is
substantially more valuable to B than to A.
The result is worse completion time for B and lower overall cluster efficiency,
even though both the cloud and the tenants would benefit from a different
allocation.

A natural refinement is to let the cloud place each application on the first
available compatible machine.
If A arrives first, it takes HW1 for its lowest cost, while B is left with HW2
at higher cost and lower performance, as in a first-come-first-served (FCFS)
setting.
This improves utilization, but the allocation is still globally suboptimal
because the cloud cannot see that HW1 has higher marginal value to B than to A.
If the cloud instead relies on spot-style preemption, B may preempt A just
before a checkpoint, as in FCFS-P, wasting work and raising overall cost.
At that moment, the operator still lacks the visibility to know that A is a
particularly bad victim to reclaim.
In both cases, beneficial exchanges are blocked because each sees only its local state.

\subsection{What the Cloud Interface Must Provide}
This example suggests three requirements for a better public-cloud interface.
First, allocations must be continuously revisitable: if tenant utility and
reconfiguration cost change over time, the cloud cannot freeze resource
decisions at launch and achieve optimal outcomes.

Second, the interface must support coordination not just between tenant
and cloud, but also among tenants.
Tenants must be able to express valuations over topology and failure domains,
not just hardware types, so that resources can move toward higher-value uses
and operators can steer exchanges using their view of shared infrastructure
constraints.
Local decision making alone is not enough when independently optimized
applications share the same substrate \cite{carbonresponder,stojkovic:tapas}.

Third, the interface must remain thin.
Centralized application logic or rich telemetry exchange across trust boundaries
is neither desirable nor feasible in the public cloud, so it needs a mechanism
that aligns decentralized decisions without exposing internal state.
That same constraint also rules out full per-resource price transparency:
tenants should see only the price information needed to reason about resources
they currently hold or can request relative to them.
The public cloud therefore needs a low-bandwidth coordination interface for
continuous renegotiation of running allocations.
\section{Design Principles: A Price for Everything}%
\label{sec:insight}
The requirements above imply three design principles for a better public-cloud
allocation interface.
\sys adopts these principles by treating price, attached to resource requests
and current ownership over a shared compatibility-and-placement resource
structure, as the shared external signal through which participants express
local decisions.

\subsection{Price as the Narrow Waist}
\label{subsec:price_narrow}
Modern cloud applications and operators both maintain rich private state:
workload phase, SLO violation cost, checkpoint timing, power headroom,
maintenance plans, and business policy.
Exchanging that state directly across trust boundaries is impractical, but the
public cloud already exposes one shared contract to all parties: price.
Public-cloud resources, however, are not flat: they are structured by
compatibility and placement, which determines which alternatives can substitute
for one another.
Tenants can therefore use price to express the current utility of acquiring or
retaining a resource within that structure, while operators use price to
express changing desirability and cost of supplying constrained portions of it.
Price thus serves as a narrow waist: a low-bandwidth signal that carries enough
information for coordination without requiring either side to reveal its
reasoning.
It also supports scoped rather than global visibility.

\begin{figure}%
\centering
\includegraphics[scale=\figscale]{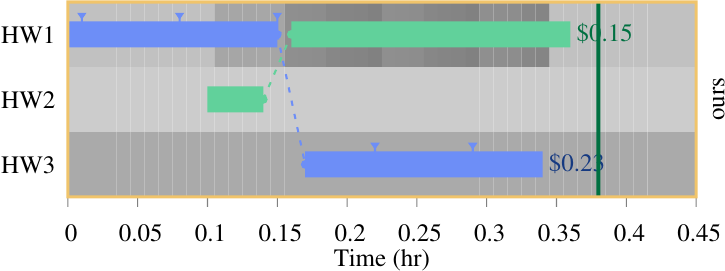}%
\caption{App A acquires HW1 first and App B acquires HW2 on dispatch, but B
raises its willingness to pay for HW1 over time. The higher price eventually
induces A to migrate after reaching a checkpoint, avoiding wasted work before
moving the resource to the tenant that values it more.}%
\label{fig:motiv_lfc}%
\end{figure}

\subsection{Continuous Renegotiation}
\label{subsec:conti_reneg}
Launch-time contracts are too rigid because both the value of a running
resource and the cost of giving it up evolve during execution.
On-demand instances freeze that decision too early, while spot-style
preemption makes allocations reclaimable but ignores when a tenant can
relinquish cheaply.
\sys instead keeps ownership continuously contestable: a running tenant keeps a
resource only while its current value exceeds competing demand.
In \autoref{fig:motiv_lfc}, this lets App A retain HW1 until a checkpoint
makes migration cheap, after which the resource can move to the tenant that
values it more.

\subsection{Decentralized Policies, Centralized Arbitration}
\label{subsec:decenti_centi}
These decisions must remain local.
Tenants should keep application-specific utility, topology preferences, and
migration logic inside their own runtimes, while operators should keep
telemetry, business policy, and pricing logic inside their control plane.
The shared system should arbitrate on externally visible signals such as
active bids, relinquishment decisions, and current ownership rather than on
application or infrastructure internals.
This preserves the public-cloud trust boundary while enabling
coordination among mutually untrusted tenants and the operator.
It also requires a resource model capturing compatibility and
placement, enabling scoped requests, scoped price visibility, and scalable
resource-level arbitration.

The resulting picture is illustrated in \autoref{fig:motiv_lfc}: App B raises
its willingness to pay for HW1 as its urgency increases, the operator could
likewise raise prices on constrained machines, and App A keeps running until
the combined price pressure no longer justifies holding the resource.
Next we show how \sys realizes these principles with resource-level
ownership, topology-aware bidding, restricted price discovery, tenant
\textsc{EconAdapters}, and operator \textsc{InfraMaps}.
\section{\sys}%
\label{sec:design}

This section describes how \sys realizes these principles in a concrete cloud
allocation system.

\begin{figure}[t]%
\centering%
\includegraphics[width=0.45\textwidth]{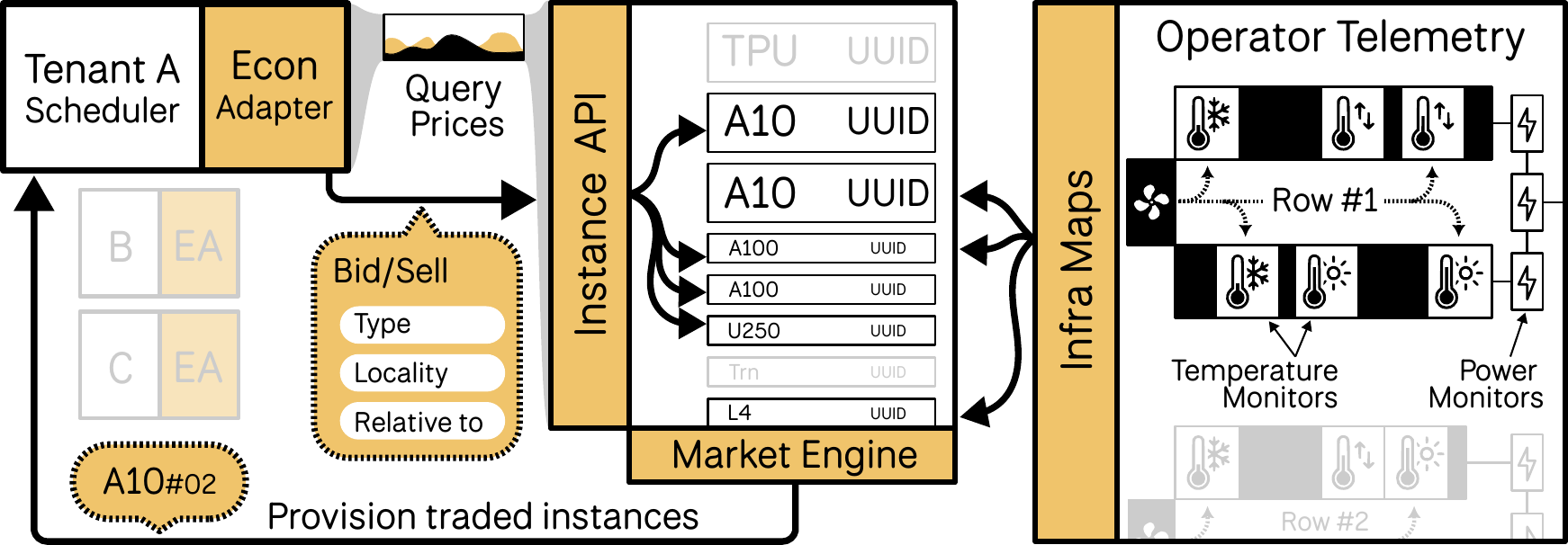}%
\caption{
    \sys overview (components highlighted).}%
\label{fig:sys}%
\end{figure}

\subsection{Overview}
\sys (\autoref{fig:sys}) exposes a continuous market for tradable compute
resources.
Each resource is a cloud instance offering, typically with a specific CPU,
memory, accelerator, and network-connectivity
configuration, although the same interface could be applied at finer
granularity.
At any moment, each resource has exactly one owner: either the operator or a
tenant.
Ownership remains contestable during execution.

Tenants interact with \sys through the instance API, while the operator
participates through automated price-setting, reclaim, and volatility-control
logic.
A participant places an active order for a scoped set of compatible resources.
Scoped price discovery can provide additional information for orders,
but is optional: applications can directly bid from local utility.
Competing active orders keep resources continuously contestable, and explicit
relinquishment or price pressure can trigger reallocation.
This section describes this ownership model, topology-aware matching structure,
restricted price discovery, and tenant- and operator-side policy modules.
Tenant \textsc{EconAdapters} generate active orders, limits, and relinquish
decisions, while operator \textsc{InfraMaps} generate price adjustments,
reclaim pressure, and volatility controls.

\subsection{Ownership, Contestability, and Billing}
Each resource instance has exactly one owner at any time.
Initially, all resources belong to the operator, which seeds the market by
offering resources at a base price.
Participants acquire resources by placing buy orders that specify a compatible
resource request.
\sys expands each request into a transactional one-cancels-others (OCO) set
of per-resource bids over the matching resource instances in the requested
scope.
Each bid enters the corresponding resource instance's local order book at the
current market price and may carry an optional limit specifying the highest
price the bidder is willing to follow if that local rate later increases.
At most one bid commits; when one does, the remaining sibling bids in the set
are canceled atomically.
A current owner may relinquish a resource explicitly, by surrendering it to the
highest-priority active matching bidder (or back to the operator's reclaim bid
if no higher tenant bid is active), or implicitly, by specifying a limit that
causes automatic relinquishment when competing demand raises the market price
above that limit.
This is how \sys keeps running allocations continuously contestable rather than
fixed at launch.

An example clarifies the control flow.
Suppose tenant A currently owns GPU $g$, while tenant B submits a buy order for
any GPU in the same NVLink domain as one it already holds.
\sys expands B's scoped request into OCO leaf bids across that subtree.
If B's demand raises $g$'s charged rate but not above A's current retention
limit, A keeps running and simply pays the higher rate while the bid remains
active.
Later, after A reaches a checkpoint and its reconfiguration cost falls, its
\textsc{EconAdapter} may lower the retention limit or explicitly relinquish
$g$.
Because B's matching bid has already been resting in the relevant leaf books,
ownership transfers immediately to B and the sibling bids in B's OCO set are
canceled atomically.
The market therefore separates two concerns that spot-style revocation
entangles: the cloud decides \emph{which} resources are contested through the
hierarchy and current prices, while each tenant decides \emph{when}
relinquishment becomes cheap enough through its own local policy.

\begin{figure}[t]%
\centering%
\includegraphics[width=0.45\textwidth]{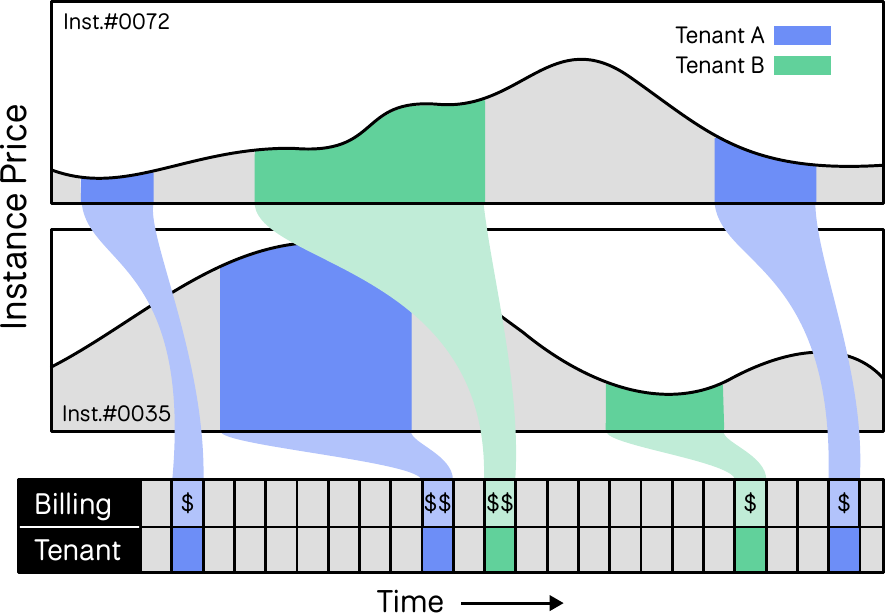}%
\caption{Resource cost equals the time integral of current price over the
interval the tenant owns a resource instance. }%
\label{fig:billing}%
\end{figure}

The operator is both the platform authority and a first-class market
participant.
Economically, it participates through the same price mechanism as tenants.
Administratively, it initializes supply, runs the control plane, and maintains
standing reclaim bids that act as per-resource floor prices.
These floor bids represent the minimum viable rate at which the operator is
willing to keep a resource allocated.
If the operator raises a floor bid for a specific resource, it increases the
charged market rate of continuing to hold that resource and can steer tenants
away from it.
As platform authority, the operator may also bound how quickly prices are
allowed to move, limiting volatility without changing the underlying bidding
semantics.

Each resource has a current charged market rate equal to the highest
active losing bid for that resource, including the operator's floor bid.
A competing order affects this rate immediately when it is placed and while it
remains active.
The current owner keeps the resource while continuing to pay that rate, but may
attach a limit indicating the highest price it is willing to tolerate for
retaining the resource.
If the charged rate rises above that limit, the owner implicitly relinquishes
the resource and the allocation is revoked immediately in the market.
Thus, when the current owner remains the highest valuer, it pays the
second-highest effective price for continuing to hold the resource.
Tenant bills are the time integral of this charged rate over the interval for
which the tenant owns the resource, as illustrated in \autoref{fig:billing}.
Crossing a limit does not introduce an additional grace period, although the
physical handoff may still take a short time (typically on the order of
10--100\,ms) while the
control plane finalizes transfer and infrastructure state.

\subsection{Market Structure: Topology-Aware Matching}
\label{sec:hiermatch}
Cloud resources are rarely fungible: the value of a resource instance depends
on both its type and its location in the cloud topology.
\sys therefore organizes the market as a forest of type-specific trees.
Each tree root corresponds to a compatible resource offering, and internal
nodes refine that offering by placement and failure-domain structure, e.g.
availability zone, rack, NVLink domain, and physical host, down to individual
resource instances.
This lets tenants request resources with locality, lets operators scope prices
to subtrees, and improves scalability because most orders touch only
one type tree and often a much smaller locality-scoped subtree.

\begin{figure}[t]%
\centering%
\includegraphics[width=0.45\textwidth]{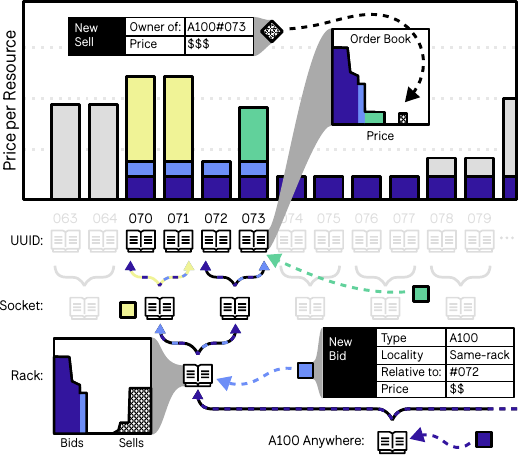}%
\caption{
    Each resource instance's order book is a leaf in one of the type-specific
    trees for the hierarchical topology. Order books on inner nodes aggregate
    orders in the books below. }%
\label{fig:hiermark}%
\end{figure}

\sys realizes this using a hierarchical matching engine.
Leaves correspond to concrete resource instances, while internal books
represent the corresponding topology groups.
Buy orders may enter at any internal book to request any satisfying descendant,
while relinquishment always occurs at leaves because ownership is defined for a
specific resource instance.
When a topology-scoped buy order arrives, its OCO bid set becomes active
throughout the relevant subtree, and each constituent bid contributes to the
charged market rate of its corresponding descendant resource.
This subtree-wide pressure is deliberate: a scoped order expresses demand for
any matching descendant without exposing per-resource prices within that scope.
If no descendant resource is relinquished, the order remains active in the
subtree so that a future explicit sell or limit crossing at any matching leaf
can satisfy it.
When a leaf resource is relinquished, the highest matching active leaf bid in
the subtree becomes the next owner, the winning bid is removed from the other
books it occupies, and transfer begins.
We show an example hierarchy with CPU-socket- and rack-level groupings in
\autoref{fig:hiermark}.

\subsection{Price Discovery and Restricted Views}
\label{sec:price_discovery}
Tenants do not require, and cloud operators may not want to expose, a global
view of every resource’s price.
Instead, \sys exposes prices through the hierarchy at the level relevant to a
tenant’s current decisions.
The instance API in \autoref{fig:sys} translates conventional cloud-style
requests into market operations over the hierarchy and exposes current prices at
the queried scope.
These prices are the rates a tenant would have to meet or exceed to acquire the
cheapest currently acquirable matching descendant resource within that scope
under current competition.
They serve as informational snapshots for shaping subsequent orders, but they
are not required for participation: the charged market rate is determined by
the active orders currently in the market.

This restriction is deliberate rather than merely a privacy concession.
A tenant rarely needs the full price surface of the cluster; it primarily needs
to compare its current allocation against nearby substitutes that differ in type
or locality.
By revealing prices only at scopes reachable from owned resources, \sys lets a
training job ask questions such as ``what would another H100 in this rack
cost?'' or ``what is the cheapest compatible GPU anywhere?'' without exposing
unrelated prices in distant subtrees.
This keeps the interface close to conventional cloud requests while still
enabling topology-sensitive exchanges.
It also reduces strategic surface area: tenants can react to relative prices
relevant to their next move, but cannot cheaply reconstruct global demand or
operator policy from a complete market tape.

\sys enforces a visible pricing domain for each tenant: the set of hierarchy
nodes whose prices that tenant may query.
At the start, a tenant can query only root scopes, e.g. “any resource of type
$X$.”
Once the tenant owns a resource, it can also query prices at ancestor nodes for
that resource, allowing it to reason about nearby or otherwise related
alternatives without revealing the entire market.
\textsc{InfraMaps} are the main exception because the operator has privileged
visibility and may price specific resources directly.
Coarse bids therefore remain simple to express, while more specific bids can
still target favored subtrees when locality matters.

\subsection{Tenant-Side Logic: EconAdapter}
\label{sec:econadapter}
\textsc{EconAdapter} is the tenant-side module that translates
application state into active orders, optional retention limits, and explicit
relinquishment decisions.
It preserves our separation of concerns: the application runtime or
autoscaler still decides when more or fewer resources would be useful, while
\textsc{EconAdapter} decides how to express those preferences in \sys.
In particular, it converts application-specific utility into bid rates for
new resources and a limit for currently owned resources.

To do so, \textsc{EconAdapter} consumes inputs that many modern autoscalers
already maintain: current utility gap, marginal utility of additional
resources, budget or penalty model, and reconfiguration overhead.
The latter is crucial, because the cost of giving up a resource often depends on
current application state, such as progress towards a checkpoint for training
jobs or warm state for inference services.
Autoscalers thus provide a natural point for integrating
\textsc{EconAdapter} without exposing application internals to the cloud.
The concrete adapters we use for training, inference, and batch analytics are
described later in \autoref{sec:eval}; here, the key point is that
\textsc{EconAdapter} turns application utility and reconfiguration cost into the
market actions required by \sys.

\subsection{Operator-Side Logic: InfraMaps}
\label{sec:infra_maps}
\textsc{InfraMaps} are the operator analogue of \textsc{EconAdapter}.
They translate operator state into the floor prices, reclaim pressure,
and volatility controls that shape the market without exposing raw telemetry or
policy details to tenants.
The operator participates through the same price mechanism as
tenants, but with broader visibility and administrative authority.

\textsc{InfraMaps} consume inputs from the operator's control plane and data
center infrastructure management (DCIM) systems, including power and cooling
headroom, maintenance plans, rack utilization, and service or business
policy.
They convert these inputs into resource-specific or subtree-specific price
adjustments that are injected into the market as standing orders.
Because the operator has privileged visibility, it can target specific resource
instances directly and can maintain reclaim bids for resources that are already
allocated.

This mechanism lets the operator steer demand using price instead of hard
intervention whenever possible.
For example, an \textsc{InfraMap} may raise the floor price of resources in a
power-constrained rack, increase reclaim pressure on resources scheduled for
maintenance, or bound the rate at which prices are allowed to rise during a
burst.
\section{Evaluation}%
\label{sec:eval}
We evaluate the following questions:
\begin{compactitem}[\labelitemi]
  \item Does \sys improve allocation efficiency under heterogeneous
    multi-tenant contention? (\autoref{sec:eval:efficiency})

  \item Does \sys improve tenant outcomes through better cost-performance
    trade-offs and topology-sensitive placements?
    (\autoref{sec:eval:tenants})

  \item Does \sys give operators a practical control lever through
    price-based steering? (\autoref{sec:eval:operators})

  \item Is \sys practical to deploy at cloud scale, with acceptable
    robustness and integration effort? (\autoref{sec:eval:practical})
\end{compactitem}

\subsection{Setup}
\paragraph{Workloads}
Our experiments are trace- and profile-driven and use representative
autoscalers rather than synthetic bidding policies.
We study three workload classes that capture much of today's accelerator
demand: LLM inference, distributed DNN training, and batch analytics.
\autoref{tab:rep_wrklds} summarizes the systems and the sources of
reconfiguration overheads, traces, and hardware profiles that instantiate
them.

Inference tenants run Nvidia Dynamo's Planner on random 200\,s windows from
Azure LLM serving traces.
Planner predicts prefill and decode node requirements from token lengths and
tenant SLOs using profiled GPU data, and we instantiate diverse inference
tenants by sampling SLO configurations from the open-source documentation.
Inference tenants bid according to the reduction in SLA penalties from
additional hardware; we use Microsoft's online-services SLA for model serving,
where
P999 and P99 latency violations incur 10\% and 25\% service credits,
respectively~\cite{msft_slaonline2026}.

We configure training and batch tenants with deadlines and derive bids from
remaining work versus remaining time, in the spirit of Uniform
Progress~\cite{wu:cantbelate}; training runs use 20 epochs.
Batch tenants run on any compatible node, while training tenants are
topology-sensitive and use Sailor profiles to derive throughput-aware bids.
Concretely, we split Sailor's dynamic-programming solver into a shadow planner
that continuously recomputes the best priced cluster configuration and a second
planner that assigns acquired hardware; the \textsc{EconAdapter} relinquishes
nodes acquired but not used by the planner.
We assign all tenants comparable budgets to limit SLO penalty spend while
preserving the objective of minimizing total cost.
We discuss implementation effort in \autoref{sec:eval:integration}.

\paragraph{Clusters and Pricing}
We evaluate right-sized, slightly oversubscribed, and heavily oversubscribed
clusters with mixtures of H100 and A100 GPUs, following the cluster-demand
regimes from Faro~\cite{faro}.
These regimes let us study both light and severe contention while holding
workload models fixed.
We focus on GPUs because both temporal reconfiguration costs and spatial
topology materially affect tenant value.
Across these regimes, we vary mixes of inference, training, and batch tenants
with different autoscaler parameters, penalties, and budgets.
Unless noted otherwise, we anchor baseline prices to comparable public-cloud
GPU prices~\cite{aws_pricing_calculator}; for \sys, we scale provider base
prices to approximate break-even at full utilization using a 70\% average
utilization assumption~\cite{kilcioglu2017usage}.

\begin{table}[t]
    \caption{Representative workloads. We use traces and performance profiles
        to derive tenant scaling and bidding policies.}
    \centering
    \footnotesize
    \begin{tabular}{p{2cm}cccc}
        \toprule
        \textbf{System} & \textbf{Reconfig. Overhead} & \textbf{Trace} & \textbf{Profile} \\
        \midrule
        \textbf{LLM Inference} \\
        Dynamo~\cite{nvdynamo}       & ~1 min~\cite{fu:serverlessllm} & \cite{qlm2024patke} & AI-Config.~\cite{dynamo:aiconfigurator}\\
        \textbf{DNN Training} \\
        Sailor~\cite{sailor} & 1-4 min~\cite{lian:up} &  N/A                 & Sailor~\cite{sailor} \\
        \textbf{Batch Analytics} \\
        Parabricks~\cite{nvidia:parabricks} & 4-12 min~\cite{wu:cantbelate} & N/A                  & PB-bench~\cite{nvidia:parabricks_benchmarks}\\
        \bottomrule
    \end{tabular}
    \label{tab:rep_wrklds}
\end{table}

\paragraph{Baselines}
We compare against two baselines that reflect today's public-cloud contracts.
\begin{compactitem}[\labelitemi]
    \item \textbf{First-come-first-served (FCFS)}: requests
    allocate in arrival order, and tenants wait if matching HW is
    occupied.
    \item \textbf{FCFS with preemption (FCFS-P)}: inference tenants may
    preempt training tenants using a spot-style model.
\end{compactitem}

To isolate the effect of the cloud interface rather than tenant logic, all
systems use the same tenant-side autoscalers, performance profiles,
SLOs/deadlines, penalty models, budgets, resource-compatibility sets, and
freeing behavior.
Applications request any hardware they can use and later relinquish excess
capacity based on lowest marginal utility.
If the cloud provisions multiple compatible instances, the application later
prunes or deallocates surplus nodes using the same marginal-utility policy.
Thus, the primary difference between \sys, FCFS, and FCFS-P is the cloud-side
allocation contract: continuous negotiation, static allocation, or spot-style
preemption.

\paragraph{Metrics}
Our primary metric is \emph{performance retention under contention}: for each
tenant, we divide the performance achieved in a multi-tenant run by the
performance achieved when that same tenant runs alone on the same cluster.
For inference, performance is the fraction of the latency objective achieved;
for training and batch, it is normalized progress toward the configured
deadline.
For each scenario, we report both the distribution of per-tenant performance
retention and the mean retention across tenants.

For tenant-side analyses, we additionally report total execution cost and
performance per cost, computed as achieved multi-tenant performance divided by
total execution cost.
We choose latency targets and deadlines so that sampled tenants are
satisfiable when they run alone in the full cluster.
We use these normalized metrics only to compare cloud interfaces across
heterogeneous workloads, not to claim that the underlying application
objectives are identical.

\subsection{Allocation Efficiency Under Contention}
\label{sec:eval:efficiency}

\begin{figure}%
\centering
\includegraphics[scale=\figscale]{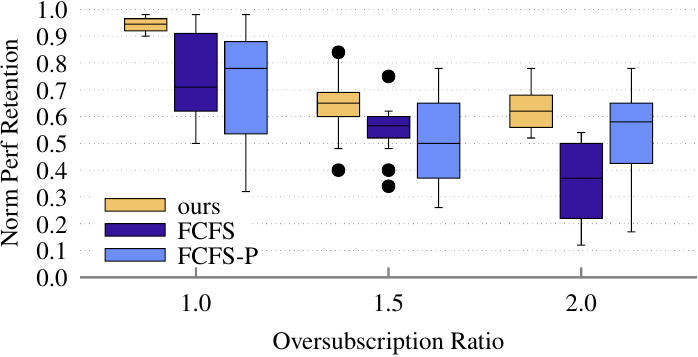}%
\caption{Performance retained for clusters of various composition.}%
\label{fig:eq1_2_subscriptionRatio}%
\end{figure}

\sys improves allocation quality across all contention regimes in
\autoref{fig:eq1_2_subscriptionRatio}.
Compared with FCFS, it reduces performance degradation by 17\% in
right-sized clusters, 8\% in slightly oversubscribed clusters, and
23\% in heavily oversubscribed clusters.
Compared with FCFS-P, it reduces degradation by 19\%,
12\%, and 8\%, respectively.
The boxplots are also generally tighter under \sys, suggesting that
continuous renegotiation reduces the number of tenants that suffer very poor
allocations under contention.

\sys gains from revisiting allocations as tenant utility and competing demand
change.
In FCFS, early tenants can hold resources that later arrivals value more, even
after those resources cease to be critical to the current owner.
Even right-sized clusters exhibit this effect because tenants may still acquire
more resources than they need to meet their current objectives.
FCFS-P partially corrects this with preemption, but it chooses victims
coarsely and can evict already underprovisioned tenants at expensive moments.
\sys instead lets tenants contest running allocations continuously and
relinquish resources when price pressure exceeds current value, which shifts
resources toward higher-value uses while avoiding wasted work.

The later case studies explain where these gains come from.
First, urgent tenants can raise bids and reclaim resources from lower-value
uses without waiting for an instance to terminate.
Second, low-urgency tenants can trade down to cheaper hardware or pause when
prices rise rather than hold expensive allocations by default.
Third, topology-sensitive tenants can express that some compatible resources
are much better than others, enabling exchanges that FCFS and FCFS-P cannot
even represent.
Together, these effects matter most under moderate and heavy contention, where
static launch-time contracts leave the most value stranded.

As expected, all three systems converge in undersubscribed settings because
contention disappears.

\subsection{Tenant Outcomes}
\label{sec:eval:tenants}

Beyond improving average allocation quality, \sys improves tenant outcomes in
two ways.
It lets tenants adjust cost-performance trade-offs online as prices and
urgency change, and it lets topology-sensitive jobs express placement
preferences that static cloud contracts cannot capture.

\subsubsection{Cost-Performance Trade-Offs}
\begin{figure}%
\centering
\includegraphics[scale=\figscale]{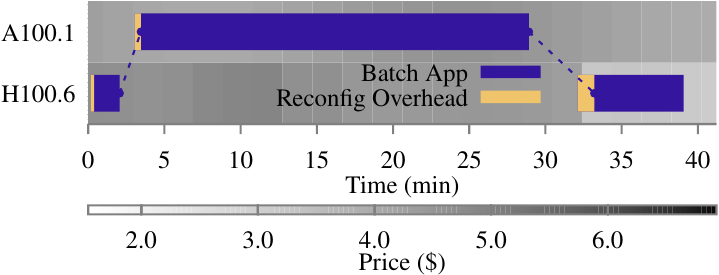}%
\caption{\sys lets tenants react to live prices and shift to cheaper allocations when urgency is low.}%
\label{fig:eq2_1_cheap}%
\end{figure}

\begin{figure}%
\centering
\includegraphics[scale=\figscale]{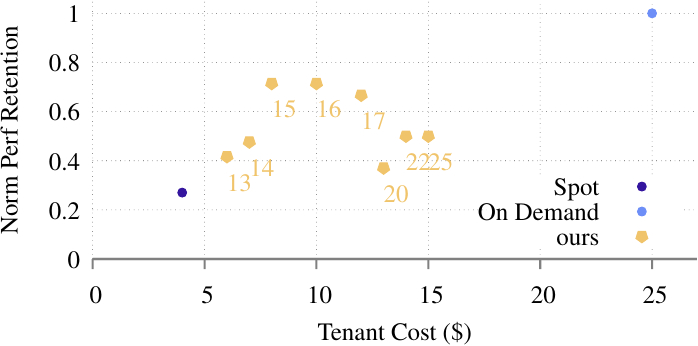}%
\caption{\sys lets tenants navigate a broad cost-performance frontier between spot-like and on-demand-like behavior. Budgets provided to the app are provided next to \sys points.}%
\label{fig:eq2_2_costEff}%
\end{figure}

\sys exposes a broad cost-performance spectrum between fixed on-demand allocations
and coarse spot preemption.
Tenants can update bids and retention limits as their remaining work,
remaining time, and current prices change, rather than commit to a single
launch-time choice.
In \autoref{fig:eq2_1_cheap}, a batch tenant reacts to an H100 price increase
by moving to a cheaper A100, pausing once it is sufficiently ahead of
schedule, and later resuming on a cheaper H100.
This behavior mirrors UniformProgress~\cite{wu:cantbelate}, but \sys lets the
tenant realize it through continuous bids instead of switching between rigid
service classes.
\autoref{fig:eq2_2_costEff} shows that different bidding strategies let tenants
span a broad range of cost-performance points.

\begin{figure}%
\centering
\includegraphics[scale=\figscale]{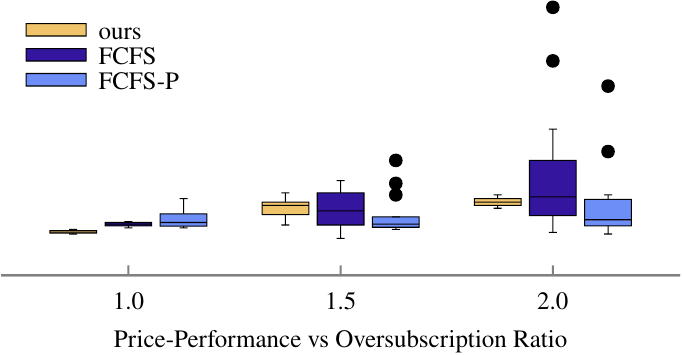}%
\caption{\sys offers more consistent performance per cost than current cloud baselines.}%
\label{fig:eq4_1_fair}%
\end{figure}

\sys also delivers more consistent performance per cost than FCFS and FCFS-P.
In FCFS, a resource delivered too late to help a tenant meet its deadline can
cost as much as the same resource delivered earlier, even though it yields less
value.
We compare performance per cost across workloads only after normalizing each
tenant's performance to its own objective: inference tenants contribute the
fraction of their latency target achieved, while training and batch tenants
contribute normalized progress toward their deadlines.
Because we calibrate SLOs, deadlines, and penalty models so that equal
shortfalls correspond to comparable loss within our setup, this metric lets us
compare how well each cloud interface converts spend into useful progress.
In \autoref{fig:eq4_1_fair}, \sys produces tighter distributions of
performance per cost across demand regimes because tenants can bid in line
with their utility curves and avoid paying full price for poorly timed,
low-value allocations.

\subsubsection{Topology-Sensitive Placements}
\begin{figure}%
\centering
\includegraphics[scale=\figscale]{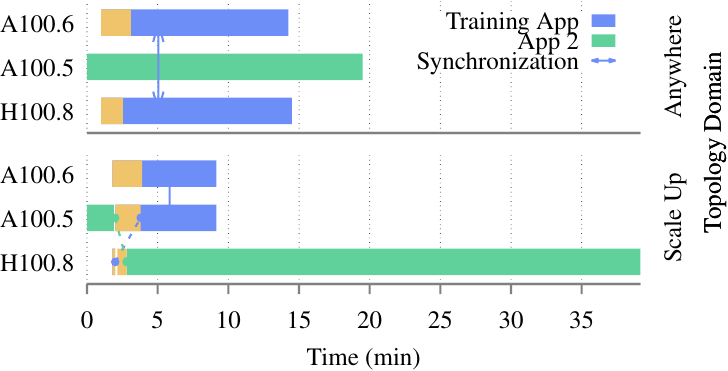}%
\caption{Topology-aware bidding lets a training job align its allocation within a favorable scale-up domain and nearly doubles performance.}%
\label{fig:eq1_4_topo}%
\end{figure}

Topology-aware bidding lets training jobs express that one compatible GPU may
be far more valuable than another because of its position in the scale-up
domain.
To isolate this effect, we run training tenants in a 1.5x oversubscribed
cluster and toggle topology-aware bidding while holding the rest of \sys
fixed.
With topology-aware bidding enabled, the training job can target a GPU in the
same scale-up domain as one it already owns and nearly doubles performance
relative to topology-oblivious bidding, as shown in
\autoref{fig:eq1_4_topo}.
This result shows that price alone is not enough: tenants must also be able to
express structured preferences over placement.

\subsection{Operator-Side Control}
\label{sec:eval:operators}

\textsc{InfraMaps} give operators a soft control lever for steering demand
without exposing raw infrastructure telemetry or selecting victims directly.
This matters because operator-side constraints such as power, cooling,
maintenance, or congestion often change faster than instance lifetimes, while
today's public-cloud contracts offer only blunt responses such as admission
control, hard preemption, or power capping.

To demonstrate this capability, we replay Google power traces for two cluster
rows, treat each row as a separate power domain, and expose four nodes from
each row through \sys.
We then implement a simple \textsc{InfraMaps} policy that raises a row's floor
prices as its power headroom shrinks.
Tenants do not see raw power telemetry; they only see the price pressure that
the policy induces on the affected resources.

\begin{figure}%
\centering
\includegraphics[scale=\figscale]{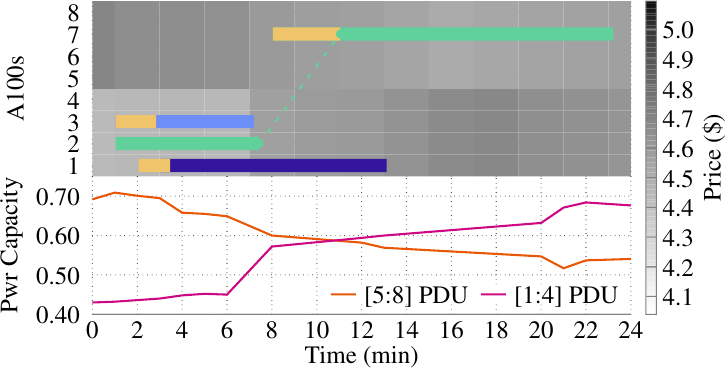}%
\caption{\sys lets an \textsc{InfraMaps} policy steer load away from a power-constrained row using prices alone.}%
\label{fig:eq3_1_pwrhmap}%
\end{figure}

\autoref{fig:eq3_1_pwrhmap} shows the resulting price-driven load shifting.
When one row becomes power-constrained, its prices rise immediately and
tenants migrate toward the less constrained row.
In the trace, the jump in row power at $t=5$ produces an immediate price
increase for that row, followed by tenant migration to the other row.
Tenants that remain on the constrained row reveal that they value those
resources enough to pay the higher rate, while less urgent tenants move away.
The operator thus reshapes load using price alone, without requiring
application-specific knowledge or direct intervention in tenant placement.
Unlike spot-style reclamation, the operator does not need to guess which tenant
would be the least costly victim; tenants self-select based on their own
current utility.

This experiment is a capability case study rather than a closed-loop power
benchmark: workload behavior does not feed back into the replayed power trace.
Still, it shows that operator-side policies can use \sys's pricing interface
as a soft-margin mechanism before resorting to disruptive actions such as power
capping or forced migration.
The same pattern applies beyond power traces: an operator could use
\textsc{InfraMaps} to steer load away from maintenance domains, thermal
hotspots, or congested subtrees while preserving the same narrow-waist
interface to tenants.

\subsection{Practicality}
\label{sec:eval:practical}

Despite the additional control-plane interaction introduced by continuous
negotiation, \sys remains practical.
We evaluate three questions: whether the market scales to cloud-sized request
rates, whether it remains effective under imperfect operating conditions, and
how much application- and operator-side logic it requires.

\subsubsection{Scalability}

\sys scales by design because it organizes resources as independent
type-specific trees (\autoref{sec:hiermatch}).
Requests for different resource types therefore proceed in parallel, and many
requests for the same type touch only a smaller locality-scoped subtree.
The central question is thus not how many resources exist in the cluster as a
whole, but how large a single type-tree can grow before the matching engine
becomes a bottleneck.

\begin{figure}%
\centering
\includegraphics[scale=\figscale]{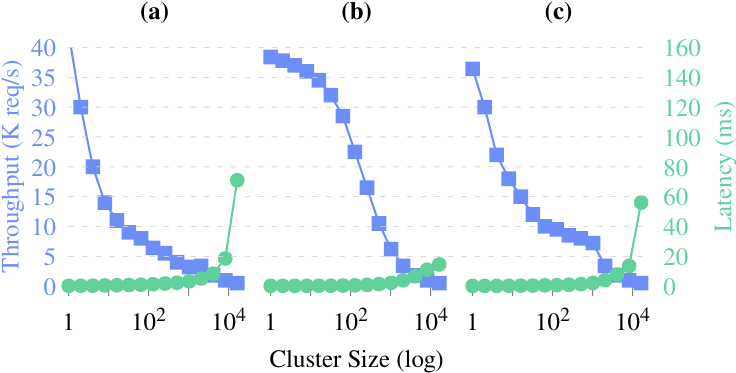}%
\caption{Scalability within a single type-tree as the number of equivalent
instances in the pool grows. \textbf{(a)} buy limit for ``anywhere''; \textbf{(b)} transfer of a
relinquished resource to the earliest queued matching buy at the finest
matching granularity; \textbf{(c)} cancel a resting buy for ``anywhere''.}%
\label{fig:mb_eq6_1_scalability}%
\end{figure}

We benchmark the heaviest operations within one type-tree as a function of tree
size, where the x-axis in \autoref{fig:mb_eq6_1_scalability} is the number of
equivalent instances in that pool.
This directly measures how many interchangeable resources \sys can manage in a
single market.
The worst case is a root-scoped buy order for ``anywhere'' because it must
remain eligible to match any descendant leaf in the tree and therefore creates
the highest degree of contention.
This case is especially expensive because resources are deprovisioned at
machine granularity, so a request for ``anywhere'' must remain eligible to
acquire any future relinquishment anywhere in the pool.

We therefore focus on three heavy operations within one tree:
first, placing a buy limit for ``anywhere'', our worst case,
second, transferring a relinquished resource to the earliest queued matching
buy order at the finest matching granularity,
and third, canceling a resting buy order for ``anywhere''.

\sys sustains up to roughly 25{,}000 requests/s while keeping request latency
below 20\,ms in these microbenchmarks.
In our end-to-end experiments, this was sufficient to support clusters of
about 10{,}000 nodes and more than 1{,}000 training and inference tenants; our
limit was available simulation capacity rather than the matching engine.
To ground these rates, training tenants submit about 3 requests/s on average
and inference tenants about 10 requests/s.
For larger and more homogeneous deployments, an operator could further shard
otherwise equivalent resources across multiple trees if a single pool became
too large, while preserving the same external interface.

\subsubsection{Robustness}
We now test whether \sys remains effective under imperfect parameter
choices and operating regimes that stress reconfiguration, price dynamics, and
tenant-side bidding logic.
For the remaining microbenchmarks, we use the slightly oversubscribed cluster
configuration from earlier.

\begin{figure}%
\centering
\includegraphics[scale=\figscale]{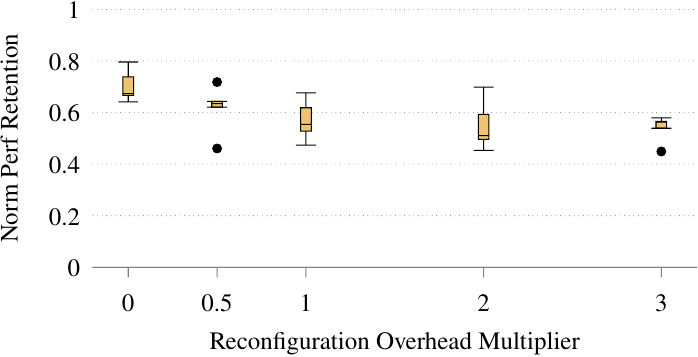}%
\caption{Lower reconfiguration overhead enables more beneficial exchanges,
while high overhead pushes \sys back toward FCFS-like behavior.}%
\label{fig:mb_eq7_1_reconfCost}%
\end{figure}

\paragraph{Reconfiguration Overhead}
Reconfiguration cost is the main counterforce to continuous renegotiation, so
we vary it by applying a uniform multiplier to all tenant overheads.
Our baseline overheads come from the representative systems in
\autoref{tab:rep_wrklds}, which reflect the current state of elastic,
fault-tolerant cloud applications; we expect these costs to fall as systems
adopt more fungible designs and cheaper reconfiguration paths.
As reconfiguration becomes more expensive, tenants switch less often because
each move wastes more paid compute time.
Since timely exchanges are the mechanism through which \sys improves
allocations, performance retention falls as this overhead rises, as shown in
\autoref{fig:mb_eq7_1_reconfCost}.
At the extreme, very high reconfiguration cost suppresses almost all exchanges
and pushes the system back toward FCFS-like behavior.
At the other extreme, cheaper reconfiguration enables more exchanges and moves
\sys closer to the ideal allocation.

\paragraph{Market Volatility}
\label{sec:eval:volatility}

\begin{figure}%
\centering
\includegraphics[scale=\figscale]{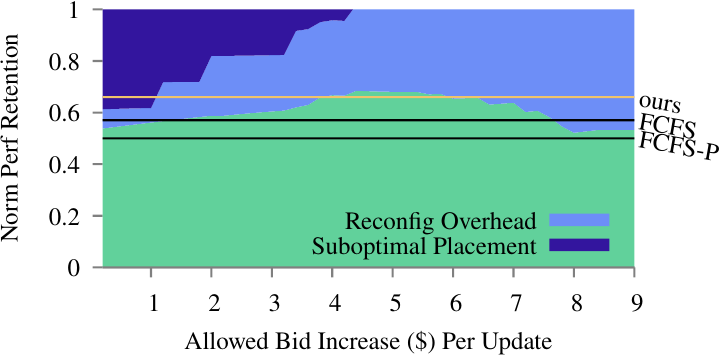}%
\caption{Excess volatility induces churn, while overly constrained prices
approach FCFS-like inefficiency; a middle ground performs best.}%
\label{fig:mb_eq8_1_volatility}%
\end{figure}

Volatility is a natural concern in a market-based allocator.
Too little price movement makes the system behave like on-demand allocation and
hides changes in tenant demand; too much movement induces churn and wastes work
through excessive reconfiguration.
\autoref{fig:mb_eq8_1_volatility} shows this trade-off.
\sys can regulate upward volatility by clipping incoming bids relative to the
current price and regulate downward volatility by bounding how quickly the
operator's floor price falls.
These controls also prevent a common failure mode around newly freed nodes:
many tenants can flood the same resource with bids, briefly spike its price,
and then leave behind unnecessary churn as those bids are canceled.
This yields a middle ground where prices move quickly enough to unlock better
allocations but not so quickly that switching overhead dominates.
The precise setting depends on workload mix, but even without tight safeguards
our experiments still outperform FCFS and FCFS-P.
In practice, operators would tune these bounds to cluster conditions.

\paragraph{Client Misconfiguration}

\begin{figure}%
\centering
\includegraphics[scale=\figscale]{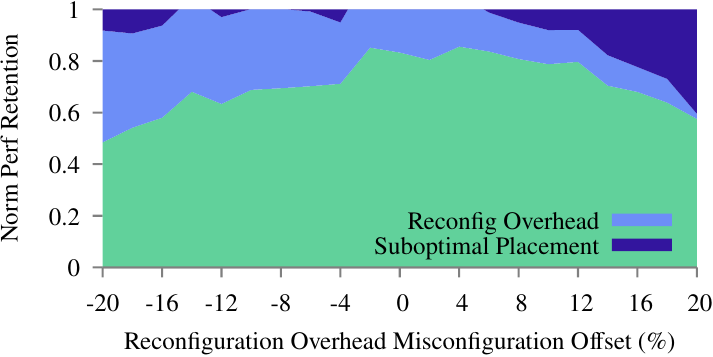}%
\caption{Underestimating reconfiguration overhead hurts more than
overestimating it, although \sys tolerates small errors of about $\pm 5\%$.}%
\label{fig:mb_eq9_1_misconfigure}%
\end{figure}

Because \textsc{EconAdapter}s price resources from local utility estimates, we
also test sensitivity to client-side misconfiguration.
We perturb only the estimated reconfiguration overhead used in bidding while
holding the true runtime overhead fixed.
Underestimating this cost hurts more than overestimating it: the tenant
reconfigures too aggressively, chases better hardware too often, and wastes
compute during each move.
Overestimation is more conservative and usually costs less as long as the
tenant starts from a reasonable placement.
At large errors, either direction can hurt performance by up to 30\%, but
\autoref{fig:mb_eq9_1_misconfigure} shows that \sys remains relatively stable
for small errors of about $\pm 5\%$.

\subsubsection{Integration Effort}
\label{sec:eval:integration}

\paragraph{Tenant Integration}
Tenant integration effort is modest.
The adapter plugs into scaling and fault-tolerance mechanisms that modern
systems already maintain.
Its hooks mirror common resilience behaviors such as checkpoint-restart,
redundancy, and shrink-and-continue.
\autoref{lst:ea_hooks} shows the core adapter logic for Sailor ML training.
The example shows how Sailor prices retained and requested GPUs from checkpoint
timing and profiled marginal utility.

\lstdefinestyle{ea-hooks}{
  float,
  captionpos=b,
  basicstyle=\ttfamily\tiny,
  numbers=left,
  numberstyle=\tiny\ttfamily,
  alsoletter={.},
  keywordstyle=[1]\color{black}\bfseries,
  keywordstyle=[2]\color{gree_cb}\bfseries,
  keywordstyle=[3]\color{blue_cb}\bfseries,
  morekeywords=[2]{APP.Profiled_marginal_utility,APP.Current_utility_gap,APP.Node_redundant,APP.Cold_start_time,APP.Time_since_chkpt,APP.Time_till_chkpt},
  morekeywords=[3]{APP.Value_per_utility_gap}
}

\begin{lstlisting}[
  style=ea-hooks,
  caption={\textsc{EconAdapter} hooks where profiling methods are marked in green and pricing implementations are denoted in blue.},
  label={lst:ea_hooks},
  firstline=1,lastline=18
]
type NodeSpec struct { // Describes the desired node to add or remove.
  NodeType string        
  Locality string        
  RelTo    string        
  Meta     map[string]any
}

// Pricing logic called by EconAdapter on every add, remove and market-update 
func Price (n EA.NodeSpec, b EA.OrderBook, gs string) float64 {
  marginal_utility := APP.Profiled_marginal_utility(n, gs)
  new_utility_gap  := APP.Current_utility_gap() - marginal_utility
  monetary_value_n := APP.Value_per_utility_gap() * new_utility_gap
  if APP.Node_redundant(n) { return monetary_val }
  var reconf_time = APP.Cold_start_time(n) // per node reconfiguration cost
  if gs == "GROW"  { reconf_time = reconf_time + APP.Time_since_chkpt(n)}
  if gs == "SHRINK"{ reconf_time = reconf_time + APP.Time_till_chkpt(n)}
  return monetary_val - (reconf_time * b.marketPrice)
}  
\end{lstlisting}

Checkpoint cadence and current application state determine the wasted work, and
hence the wasted spend, of moving between checkpoints.
The hooks are extensible: for example,
\texttt{APP.Value\_per\_utility\_gap()} can encode either SLO penalties or the
value of additional throughput.
\autoref{tab:loc_econadapter} summarizes the application-specific pricing and
profiling code we added for each workload.
For Sailor, integration requires launching a shadow DP solver and querying it
periodically for profiled marginal utility.

\paragraph{Operator Integration}
Operator integration is simpler still.
\textsc{InfraMaps} reuse the same basic structure but replace
fault-tolerance hooks with telemetry-to-price mappings.
For the power-aware policy in \autoref{fig:eq3_1_pwrhmap}, three lines of code
map power headroom to a proportional price adjustment.
Adding further operator signals amounts to adding another weighted adjustment
and rebalancing the composition.

\begin{table}[t]
  \caption{Implementation effort for integrating different prior tolerance
    approaches in the \sys \textsc{EconAdapter}.} 
  \centering
  \footnotesize
  \begin{tabular}{lcccc}
    \toprule
    \textbf{System} &
    \textbf{Fault Tol.} &
    \multicolumn{2}{c}{\textbf{\textsc{EconAdapter} (LoC)}}  \\
    \cmidrule(lr){3-4}
    & \textbf{Method} & \textbf{Price} & \textbf{Profile} & \\
    \midrule
    Dynamo LLM inference & $\blacktriangle\,\blacklozenge$  & 17 & 55  \\
    Sailor ML training   & $\blacksquare$                   & 23 & 34  \\
    Parabricks Batch Genomics  & $\blacksquare\,\blacklozenge$    & 12 & 17  \\
    \textsc{InfraMaps}: Power Traces  &   & 8 & 5  \\
    \bottomrule
  \end{tabular}

  \vspace{2pt}
  {\footnotesize
  \textbf{Legend:} $\blacktriangle$ = redundancy;\;
  $\blacksquare$ = checkpoint-restart;\;
  $\blacklozenge$ = shrink-and-continue.}

  \label{tab:loc_econadapter}
\end{table}
\section{Related Work}
\label{sec:related}

\sys sits at the intersection of shared-cluster resource management,
market-based cloud allocation, application runtimes, and operator-facing
infrastructure control.

\paragraph{Shared-Cluster Schedulers and Fleet Managers}
Shared-cluster systems expose thin allocation interfaces within one
administrative domain.
They range from fair-sharing and two-level schedulers
\cite{hindman:mesos,ghodsi:drf} to general-purpose controllers that rely on
broad visibility into workload and infrastructure state
\cite{eva,stratus,cilantro,htas}.
Large private clouds unify diverse workloads under shared fleets
and coordinating schedulers \cite{central_fleet,flux,bytedance:godel,twine},
while workload-specific managers optimize particular services or devices
\cite{222611,stojkovic:tapas,stojkovic:dynamollm}.
These systems assume common control or rich metric sharing within one domain.
\sys addresses a different setting: the public-cloud trust boundary.
It keeps utility and operator policies local to tenants and the cloud
provider, and uses price rather than shared internal state as the coordination
interface.

\paragraph{Market-Based, Spot, and Preemptible Allocation}
Systems researchers have long asked when markets help resource allocation
\cite{shneid:whynomark}. Prior work uses markets, priorities, or fair sharing
to arbitrate among tenants
\cite{vuppalapati:karma,grandl:carbyne,chaudhary:gandivafair,xchange,10.1145/3620665.3640375,sf_compute,themis,shockwave}.
Most of these systems either restrict what participants can bid on, for
example to application metrics
\cite{chaudhary:gandivafair,xchange,10.1145/3620665.3640375}, or clear
allocation decisions in fixed epochs \cite{sf_compute,themis,shockwave}.
The closest deployed cloud analogue is spot or preemptible capacity, along
with systems that improve or leverage it \cite{iqbal:cospot,yang:skypilot}.
But spot still fixes bids at launch and leaves revocation unilateral.
\sys differs in three ways: it keeps running allocations continuously
contestable, supports topology-aware bidding and operator price steering,
and does so across mutually untrusted tenants and the operator.

\paragraph{Application Runtimes and Elastic Planners}
Another line of work makes applications more fungible under today's cloud
contracts, using prediction, migration, checkpointing, or elastic planners to
tolerate changing availability
\cite{wu:cantbelate,parcae,tenplex,sailor,subramanya:sia,zeus,proteus}.
These systems adapt the application to on-demand and spot semantics rather than
changing the interface itself.
We build on this line through \textsc{EconAdapters}: modern planners already
estimate marginal utility, deadlines, and reconfiguration cost, which \sys
converts into bids and retention limits.
The key difference is again at the interface boundary.
\sys lets such planners coordinate through price with other tenants and with
the operator, rather than react independently to a static cloud contract.

\paragraph{Operator-Facing Infrastructure Control}
Another adjacent line of work exposes or exploits infrastructure signals such
as topology, power, thermal state, or energy availability
\cite{bazzaz:torhotspot,stojkovic:tapas,ecovisor,faro,carbonresponder}.
These systems highlight that operators need workload-aware levers when
infrastructure constraints shift.
\sys is complementary: rather than export raw
telemetry or centrally optimize every workload, it lets operators express such
constraints through price and lets tenants decide how to respond locally.
This operator-participation model is central to our design: the cloud is not
just another bidder, but the market designer and the actor that injects
infrastructure policy into the negotiation loop.

Finally, our choice of per-instance resources and topology trees is orthogonal
to the interface itself.
Lower-level sharing systems such as Orion optimize utilization within a single
GPU rather than negotiate ownership across machines or tenants
\cite{strati:orion}.
Many systems explore the scalability trade-offs of hierarchical or
decentralized matching and scheduling
\cite{meta:ras,shepherd,sparrow,borg}; \sys treats those mechanisms as
implementation choices beneath a continuous-negotiation interface.
\section{Discussion, Risks \& Limitations}%
\label{sec:discussion}
This section discusses adoption paths, open economic questions, and the limits
of price-based control beyond this paper.

\paragraph{Adoption Path}
\sys does not require every application to become a sophisticated trading
agent.
Simple policies can emulate today's on-demand and spot behaviors: an
on-demand-like tenant maintains a fixed allocation with high retention limits,
while a spot-like tenant uses low limits and relinquishes aggressively under
price pressure.
These policies show that \sys subsumes today's service classes
rather than replacing them with an all-or-nothing contract.
Providers could therefore start with managed policy templates, then let more
capable tenants opt into utility- and topology-aware strategies.
Emerging cloud-resource marketplaces already suggest such a migration path by
offering automated trading policies that preserve fixed-footprint
behavior~\cite{sf_compute}.
This path is practical because many modern autoscalers already expose the
inputs \textsc{EconAdapter} needs, including utility gaps, budget models, and
reconfiguration state.

\paragraph{Market Regulation}
Asymmetric visibility means the operator is not just another bidder.
It is also the market designer: it sets floor prices, volatility bounds,
admission policy, and responses to pathological behavior.
That role is deliberate.
Clouds already shape prices and enforce safety constraints; \sys makes those
powers explicit inside the allocation contract.
In practice, an operator could combine clipped bid updates, minimum holding
times, or fees on sharp price swings to damp churn during bursts, analogous to
limit-up / limit-down controls in financial markets~\cite{luldplan}.
These mechanisms need not make the market strategy-proof.
They only need to keep prices stable enough to remain a useful coordination
signal.

\paragraph{Open Economic Questions}
\sys is a systems interface, not a complete economic theory of cloud markets.
Restricted price discovery (\autoref{sec:price_discovery}) trades global market
transparency for scoped visibility, but we do not quantify the privacy or
strategic consequences of that choice.
Continuous bidding creates room for strategy, and operator volatility controls
shape how quickly prices respond to demand.
Tenants may bias bids toward the current market rate rather than reveal their
true value, especially when visible prices leak urgency.
That would still occur against external anchors such as competing
clouds' rates for comparable hardware~\cite{irwin:price}.
Our claim does not require truthful revelation.
Bids and retention limits only need to track local utility monotonically enough
to communicate relative urgency better than static contracts do.
Even biased bids can still convey useful ordinal information, for example that
a deadline-sensitive inference job currently values a GPU more than a training
job between checkpoints.
Our evaluation studies plausible tenant and operator policies under these
rules, but equilibrium analysis, incentive compatibility, and formal privacy
guarantees remain future work.

\paragraph{Limits of Price-Based Control}
Price is a soft control mechanism, not a substitute for hard operator
intervention.
Its effectiveness depends on tenant flexibility and on parameters such as
reconfiguration cost, volatility bounds, and bid calibration, as reflected in
\autoref{sec:eval:practical}.
When tenants respond too slowly, misprice resources, or infrastructure
constraints become urgent, operators may still need power capping,
throttling, or forced migration.
\sys is therefore best viewed as a first coordination layer that absorbs many
changes in demand and infrastructure state before the cloud resorts to more
disruptive mechanisms.
It adds a soft-margin control layer ahead of the hard-margin mechanisms that
operators must still retain for safety and correctness.
\section{Conclusion}
\label{sec:conclusion}

Today's public-cloud contracts fix too much too early.
This paper shows price can serve as the thin coordination interface for
continuously renegotiating running allocations under public-cloud trust
boundaries.
\sys realizes this idea as a topology-aware market with contestable resource
ownership over a shared compatibility-and-placement structure, restricted price
discovery, and local tenant and operator policy modules.
Our evaluation shows this design improves allocation quality under
contention, benefits topology-\-sen\-si\-tive workloads, gives operators a practical
steering lever, and scales to large clusters.
Public clouds should move beyond static service classes toward interfaces that
support continuous renegotiation of running allocations.
\if \ANON 0
\fi

\bibliographystyle{plain}
\bibliography{paper,bibdb/papers,bibdb/strings,bibdb/defs}
\clearpage
\appendix

\label{page:last}
\end{document}